\documentclass[
    ,final            
  ]
  {aipproc}

\layoutstyle{6x9}

\begin{document}

\title{Quantum dynamics of a vibrational mode of a membrane within an optical cavity}
\classification{42.50.Lc, 05.40.Jc, 03.67.Mn}
\keywords      {Mechanical effects of light, optical bistability, ground state cooling, entanglement}

\author{M. Karuza}{
  address={School of Science and Technology, Physics Division, University of Camerino, Italy \\ INFN, Sezione di Perugia, Italy}
}

\author{C. Biancofiore}{
  address={School of Science and Technology, Physics Division, University of Camerino, Italy \\ INFN, Sezione di Perugia, Italy}
}

\author{M. Galassi}{
  address={School of Science and Technology, Physics Division, University of Camerino, Italy \\ INFN, Sezione di Perugia, Italy}
}

\author{R. Natali}{
  address={School of Science and Technology, Physics Division, University of Camerino, Italy \\ INFN, Sezione di Perugia, Italy}
}

\author{G. Di Giuseppe}{
  address={School of Science and Technology, Physics Division, University of Camerino, Italy \\ INFN, Sezione di Perugia, Italy}
}

\author{P. Tombesi}{
  address={School of Science and Technology, Physics Division, University of Camerino, Italy \\ INFN, Sezione di Perugia, Italy}
}

\author{D. Vitali}{
  address={School of Science and Technology, Physics Division, University of Camerino, Italy \\ INFN, Sezione di Perugia, Italy}
}

\begin{abstract}
 Optomechanical systems are a promising candidate for the implementation of quantum interfaces for storing and redistributing quantum information. Here we focus on the case of a high-finesse optical cavity with a thin vibrating semitransparent membrane in the middle. We show that robust and stationary optomechanical entanglement could be achieved in the system, even in the presence of nonnegligible optical absorption in the membrane. We also present some preliminary experimental data showing radiation-pressure induced optical bistability.
\end{abstract}

\maketitle


The realization of efficient quantum networks requires the presence of quantum interfaces able to store, retrieve and redistribute quantum information. Some demonstrations have been realized with atomic ensembles, able to store both binary \cite{kimble} and continuous variable (CV) information \cite{polzik}. Another candidate, which could be well integrated with both atomic and solid-state devices, is represented by optomechanical systems, in which information can be encoded in a vibrational degree of freedom of a resonator.

Various cavity optomechanical systems have been recently developed \cite{kippenberg,amo}, but we have focused here, both theoretically and experimentally, on the setup formed by a standard Fabry-Perot cavity with a thin semitransparent membrane placed at its center representing the mechanical element \cite{harris}. A given cavity mode with annihilation operator $a$ ($[a,a^{\dag }]=1$) is driven by an intense laser with frequency $\omega_l$ and excites with its radiation pressure several vibrational modes of the membrane. However, a single mechanical mode can be considered when a bandpass filter in the detection scheme, centered around an isolated mechanical resonance, is used, and coupling between the different vibrational modes can be neglected. The system can be then described by two harmonic oscillators with Hamiltonian
\begin{equation}
H=\frac{\hbar \omega _{m}}{2}(p^{2}+q^{2})+\hbar \omega _{c}[z(q)]a^{\dagger }a+i\hbar E(a^{\dagger }e^{-i\omega
_{l}t}-ae^{i\omega _{l}t}),  \label{Ham}
\end{equation}
where $\omega _{m}$ is the frequency of the selected mechanical mode of the membrane, $q$ and $p$ ($[q,p]=i$) are the dimensionless position and momentum operators associated with it, $E$ is related to the input laser power $\mathcal{P}$ and the mode bandwidth without the membrane $\kappa_0$ by $|E|=\sqrt{2\mathcal{P}\kappa_0 /\hbar \omega _{l}}$, and $\omega _{c}[z(q)]$ is the frequency of the driven cavity mode. The latter depends upon the membrane position along the cavity axis $z(q)$ which in turn depends upon the coordinate $q$ of the membrane mode because one can write $z(q) = z_0+x_0 q$, where $z_0$ is the membrane center-of-mass position along the cavity axis and $x_0=\sqrt{\hbar/m \omega_m}$, with $m$ the effective mass of the mechanical mode \cite{harris,kimble2}.
This position dependence $\omega _{c}[z(q)]$ implicitly contains the radiation pressure interaction and can be obtained by solving the wave equation for the optical field within the cavity in the presence of the membrane with thickness $L_d$ and index of refraction $n_M$. In the case of a symmetric cavity with equal mirrors and length $L$, and if the membrane is placed close to the cavity waist at the cavity center ($z=0$), the frequency can be written as
\begin{equation}\label{eq:freq-q}
    \omega_{c}(z)=\omega_b+\frac{c}{L}\arcsin
    \left[\frac{(-1)^p\cos(2k_0z)(n_M^2-1)}{\sqrt{4 n_M^2\cot^2(n_M k_0 L_d)+(n_M^2+1)^2}}\right],
\end{equation}
where $\omega_b$ does not depend upon $z$ and its explicit expression is not relevant here, $k_0=\omega_0/c$ is the mode wave-vector without the membrane, and $p$ is the longitudinal mode index, i.e., the number of half-wavelengths within the cavity.

\subsection{Langevin equation description}

The dynamics are determined not only by the Hamiltonian of Eq.~(\ref{Ham}), but also by the fluctuation-dissipation processes affecting both the optical and the mechanical mode.
The mechanical mode is affected by a viscous force with damping rate $\gamma _{m}$ and by a Brownian stochastic force with zero mean value $\xi
(t) $, whose correlation function, in the limit of high mechanical
quality factor $Q_m=\omega_m/\gamma_m \gg 1$, can be written as \cite{GIOV01,gard}
\begin{equation}
\label{browncorre2}\left\langle \xi(t)\xi(t^{\prime})\right\rangle \simeq \gamma_{m}\left[  (2n_{0}+1) \delta(t-t^{\prime})+i
\frac{\delta^{\prime}(t-t^{\prime})}{\omega_{m}}\right]  ,
\end{equation}
where $n_{0}=\left( \exp \{\hbar \omega _{m}/k_{B}T_{0}\}-1\right) ^{-1}$ is the mean thermal excitation number at the membrane reservoir temperature
$T_{0}$, and $\delta^{\prime}(t-t^{\prime})$ denotes the derivative of the Dirac delta.

Then, due to the nonzero transmission of the cavity mirrors, the cavity field decays at rate $\kappa_0 $ and is affected by the vacuum radiation input noise $a_0^{in}(t)$, whose
only nonzero correlation functions is given by \cite{gard}
$
\langle a_0^{in}(t)a_0^{in,\dag }(t^{\prime })\rangle = \delta (t-t^{\prime })$.
The presence of the membrane and of its non-negligible optical absorption, described by the imaginary part of the refraction index, $n_M=n_R+in_I$, provides an additional loss channel for the cavity photons, together with the associated vacuum optical input noise $a_1^{in}(t)$. This is a nonlinear dissipative process which affects also the mechanical mode, because the photon absorption rate by the membrane depends upon the resonator position $q$ according to $\kappa_1[z(q)]={\rm Im}\left\{\omega_c[z(q)]\right\}$. In particular, this process is responsible for an additional stochastic force on the resonator, describing membrane heating due to absorption. Adding all these damping and noise terms to the Heisenberg equations of motion associated with the Hamiltonian of Eq.~(\ref{Ham}), one gets the following set of nonlinear quantum Langevin equations (QLE), written in
a frame rotating at $\omega _{l}$,
\begin{eqnarray}
\dot{q}& =& \omega _{m}p, \label{nonlinlang1}\\
\dot{p}& = & -\omega _{m}q-\gamma _{m}p-\partial_q \omega_c[z(q)] a^{\dag }a+\xi +i\frac{\partial_q \kappa_1[z(q)]}{\sqrt {2\kappa_1[z(q)]}}\left[a^{\dagger}a_1^{in}-a a_1^{in,\dagger}\right], \label{nonlinlang2}\\
\dot{a}& = & -i\left[\omega_c[z(q)]-\omega_l\right]a-\left[\kappa_0+\kappa_1[z(q)]\right]a+E+\sqrt{2\kappa_0}a_0^{in}+\sqrt{2\kappa_1[z(q)]}a_1^{in}, \label{nonlinlang3}
\end{eqnarray}
where $\partial_q$ denotes the derivative with respect to $q$. In order to reach the regime in which quantum effects of the optomechanical system are visible one needs a strong radiation pressure interaction and this requires an intense intracavity field, which is achievable with a large cavity finesse ${\mathcal F}$ and with enough driving power. In this case the operation point is given by a classical steady state characterized by a coherent intracavity field with amplitude $\alpha_s$ ($|\alpha_s| \gg 1$) and a deformed membrane with a new stationary position $q_s$, satisfying the conditions
\begin{eqnarray}
\label{eq:stat1}
  q_s &=& -\frac{\partial_q \omega_c[z(q_s)]|\alpha_s|^2}{\omega_m} \\
  |\alpha_s|^2 &=& \frac{|E|^2}{ \left[\kappa_0+\kappa_1[z(q_s)]\right]^2+\left[\omega_l-\omega_c[z(q_s)]\right]^2}. \label{eq:stat2}
\end{eqnarray}
Eq.~(\ref{eq:stat2}) is a nonlinear equation determining $|\alpha _{s}|$, due to the dependence of $q_s$ upon $|\alpha _{s}|$ itself given by Eq.~(\ref{eq:stat1}), and which may show optical bistability, i.e., the presence of two simultaneous stable solutions above a given threshold for the input power ${\mathcal P}$, as demonstrated in \cite{dorsel} and also with Bose-Einstein condensates \cite{brennecke}.
We have verified experimentally the presence of radiation pressure-induced optical bistability in the membrane-in-the-middle scheme by considering a symmetric Fabry-Perot cavity with length $L=9$ cm, waist $w_0=130$ $\mu$m, driven at $\lambda=1064$ nm by a Nd:YAG laser with input power ${\mathcal P}=30$ mW, and inserting at its center a macroscopic SiN membrane with thickness $L_d=500$ nm and side length $3$ mm. Optical bistability due to radiation pressure has been observed by scanning the driving laser frequency $\omega_l$ across the resonance while recording the cavity transmission on a InGaAs photodiode, either from the red and from the blue side of the cavity resonance (see Fig.~1). An evident hysteretic cycle is observed, and the data are consistent with a cavity finesse ${\mathcal F}=8000$ and an effective optomechanical coupling $ \partial_q \omega_c[z(q_s)] \sim 6$ Hz.

\begin{figure}[h]
  \includegraphics[height=.45\textwidth]{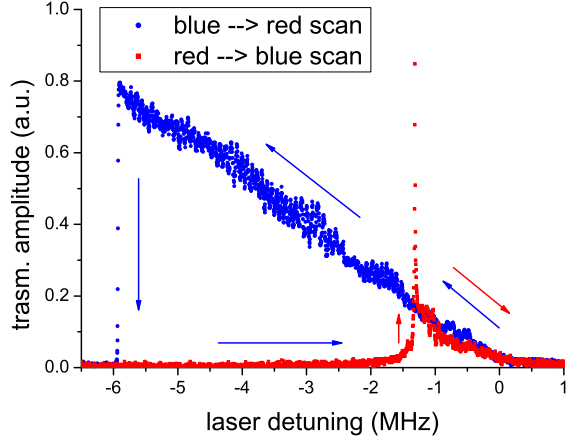}
  \caption{Plot of cavity transmission for two laser frequency scans, either from the red (red curve) and from the blue (blue curve) side of the cavity resonance relative to a TEM$_{00}$ mode.}
\end{figure}

\subsection{Quantum dynamics of the fluctuations}

The quantum behavior of the optomechanical systems could be detected in the dynamics of the fluctuations around the classical steady-state described above.
Rewriting each Heisenberg operator of Eqs.~(\ref{nonlinlang1})-(\ref{nonlinlang3}) as the classical steady state value plus an additional
fluctuation operator with zero mean value, and neglecting nonlinear terms in the equations (which is justified whenever $|\alpha _{s}|\gg 1$), one gets the following linearized QLE for the fluctuations
\begin{eqnarray}
\delta \dot{q}& =&\omega _{m}\delta p, \label{lle1}\\
\delta \dot{p}& =&-\left[\omega _{m}+\partial_q^2 \omega_c[z(q_s)]|\alpha_s|^2\right]\delta q-\gamma _{m}\delta p+G\delta X+\xi +\frac{\partial_q \kappa_1[z(q_s)]\alpha_s}{\sqrt {\kappa_1[z(q_s)]}}Y_1^{in}, \label{lle2}\\
\delta \dot{X}& =&-\left[\kappa_0+\kappa_1[z(q_s)]\right] \delta X+\Delta \delta Y-\sqrt{2}\partial_q \kappa_1[z(q_s)]\alpha_s \delta q \nonumber \\
&& +
\sqrt{2\kappa_0}X_0^{in}+\sqrt{2\kappa_1[z(q_s)]}X_1^{in}, \label{lle3}\\
\delta \dot{Y}& =&-\left[\kappa_0+\kappa_1[z(q_s)]\right] \delta Y-\Delta \delta X+G\delta q+\sqrt{2\kappa_0}Y_0^{in}+\sqrt{2\kappa_1[z(q_s)]}Y_1^{in}. \label{lle4}
\end{eqnarray}
We have chosen the phase reference of the cavity field so that $\alpha _{s}$ is real and positive, and we have defined the detuning $\Delta = \omega_c[z(q_s)]-\omega_l$, the cavity field
quadratures $\delta X\equiv (\delta a+\delta a^{\dag })/\sqrt{2}$ and $\delta Y\equiv (\delta a-\delta a^{\dag })/i\sqrt{2}$ and the
corresponding Hermitian input
noise operators $X_j^{in}\equiv (a_j^{in}+a_j^{in,\dag })/\sqrt{2}$ and $%
Y_j^{in}\equiv (a_j^{in}-a_j^{in,\dag })/i\sqrt{2}$, $j=0,1$. The linearized QLE show that the mechanical mode is coupled to the cavity mode quadrature
fluctuations by the effective optomechanical coupling
\begin{equation}
G=-\partial_q \omega_c[z(q_s)]\alpha _{s}\sqrt{2}=-2\partial_q \omega_c[z(q_s)]\sqrt{\frac{\mathcal{P}%
\kappa_0 }{\hbar\omega _{l}\left[\left[\kappa_0+\kappa_1[z(q_s)]\right]^{2}+\Delta ^{2}\right] }}, \label{optoc}
\end{equation}%
which can be made very large by increasing the intracavity amplitude $\alpha _{s}$. In addition to cavity decay and noise terms related to $\kappa_1[z(q_s)]$, optical absorption by the membrane is responsible also for two additional terms in the linearized QLE of Eqs.~(\ref{lle1})-(\ref{lle4}) which are usually neglected in standard treatments: the noise term proportional to $Y_1^{in}$ describing heating of the mechanical resonator, and the dissipative coupling term $\sqrt{2}\partial_q \kappa_1[z(q_s)]\alpha_s \delta q$.
It is therefore interesting to verify if and when absorption by the membrane may hinder achieving the quantum regime for the optomechanical system under investigation. In fact, it has been already shown that such a system may achieve a stationary state showing simultaneously both ground state cooling of the mechanical element and robust optomechanical entanglement \cite{prl07,output}. The latter could be extremely useful for quantum information applications because, due to its stationary nature, it is robust to decoherence and available at any time.

The stationary solution of Eqs.~(\ref{lle1})-(\ref{lle4}) can be obtained with standard techniques and it is reached provided that the system is stable. Stability conditions can be derived from the Routh-Hurwitz criterion \cite{output}, and are slightly different from those discussed in \cite{prl07,output} due to the new terms in the linearized QLE. Stability will be discussed in more detail elsewhere and will be assumed to be satisfied from now on. Due to the linearization and to the Gaussian nature of the noise operators, the steady state is a zero-mean Gaussian state, completely
characterized by its covariance matrix (CM). The latter is given by the $
4\times 4$ matrix $V$ with elements $
V_{lm}=\left\langle u_{l}\left( \infty \right) u_{m}\left( \infty \right) +u_{m}\left( \infty \right) u_{l}\left( \infty \right)
\right\rangle /2$, where $u_{m}(\infty)$ is the asymptotic value of the $m$-th component of the vector of quadrature fluctuations
$ u(t)=\left( \delta q(t),\delta p(t),\delta X(t),\delta Y(t)\right) ^{T }$. The steady state CM can be determined by solving the Lyapunov equation \cite{prl07,output}
\begin{equation}
A V+V A^{T}=-D,  \label{Lyapunov}
\end{equation}%
with $A$ the drift matrix
\begin{equation}
A=\left(
\begin{array}{cccc}
0 & \omega _{m} & 0 & 0 \\
-\omega _{m} -\partial_q^2 \omega_c[z(q_s)]|\alpha_s|^2 & -\gamma _{m} & G & 0 \\
-\sqrt{2}\partial_q \kappa_1[z(q_s)]\alpha_s & 0 & -\kappa_T[z(q_s)]  & \Delta  \\
G & 0 & -\Delta  & -\kappa_T[z(q_s)]
\end{array}%
\right) ,
\end{equation}%
where $\kappa_T[z(q_s)]=\kappa_0+\kappa_1[z(q_s)]$ is the total cavity decay rate, and $D$ the $4\times 4$ diffusion matrix
\begin{equation}
D=\left(
\begin{array}{cccc}
0 & 0 & 0 & 0 \\
0 & \gamma _{m}\left( 2n_{0}+1\right) + \frac{[\partial_q \kappa_1[z(q_s)]]^2|\alpha_s|^2}{2\kappa_1[z(q_s)]}& 0 & \frac{\partial_q \kappa_1[z(q_s)] \alpha_s}{\sqrt{2}} \\
0 & 0 & \kappa_T[z(q_s)]  & 0  \\
0 & \frac{\partial_q \kappa_1[z(q_s)] \alpha_s}{\sqrt{2}} & 0  & \kappa_T[z(q_s)]
\end{array}%
\right) .
\end{equation}%
Equation~(\ref{Lyapunov}) is a linear equation for the CM $V$ which can be solved, but its solution is very cumbersome and will not be reported here.
The CM contains all the informations about the steady state: in particular, the mean energy of the mechanical resonator is given by
$U=\hbar \omega _{m}\left[ \left\langle \delta q^{2}\right\rangle +\left\langle \delta p^{2}\right\rangle \right]/2 = \hbar \omega_{m}\left[ V_{11} +V_{22} \right]/2 \equiv \hbar \omega _{m}\left( n+1/2\right)$,
where $n$ is the effective mean vibrational number of the resonator. Obviously, in the
absence of radiation pressure coupling it is $n=n_{0}$, where $n_{0}$ corresponds to the actual temperature of the environment $T_{0}$. The
optomechanical coupling with the cavity mode can be used to engineer an effective bath of much lower temperature $T\ll T_{0}$, so that the
mechanical resonator is cooled. Moreover from $V$ one can also calculate the optomechanical entanglement between the cavity mode and the resonator at the steady state. We adopt as entanglement measure the logarithmic negativity
$E_{N}$, which is a convenient entanglement measure, easy to compute and also
additive. It is defined as \cite{logneg} $
E_{N}=\max [0,-\ln 2\eta ^{-}]
$, where $\eta ^{-}\equiv 2^{-1/2}\left[ \Sigma (V)-\left[ \Sigma (%
V)^{2}-4\det V\right] ^{1/2}\right] ^{1/2}$and $\Sigma (V)\equiv \det V_{1}+\det V_{2}-2\det V_{c}$, with $V_{1},V_{2}$ and $V_{c}$ being the $2\times 2$ sub-block matrices of $V$
\begin{equation}
V\equiv \left(
\begin{array}{cc}
V_{1} & V_{c} \\
V_{c}^{T} & V_{2}%
\end{array}%
\right) .  \label{CMatrix}
\end{equation}%
Ref.~\cite{output} has showed that stationary ground state cooling of the mechanical resonator and optomechanical entanglement could be simultaneously achieved and experimental results in this direction have been recently attained \cite{groblacher}. Choosing parameters comparable to those of Ref.~\cite{kimble2} in the linearized QLE above, we see that simultaneous cooling and entanglement is possible also in the membrane-in-the-middle scheme, because membrane absorption does not represent a serious limitation. This is shown in Fig.~2 where both $n$ and $E_{N}$ are plotted versus temperature (a) and versus the input power (b). We see that, in the absence of technical limitations, ground state cooling and entanglement could be achieved even at room temperature.

\begin{figure}
  \includegraphics[width=.99\textwidth]{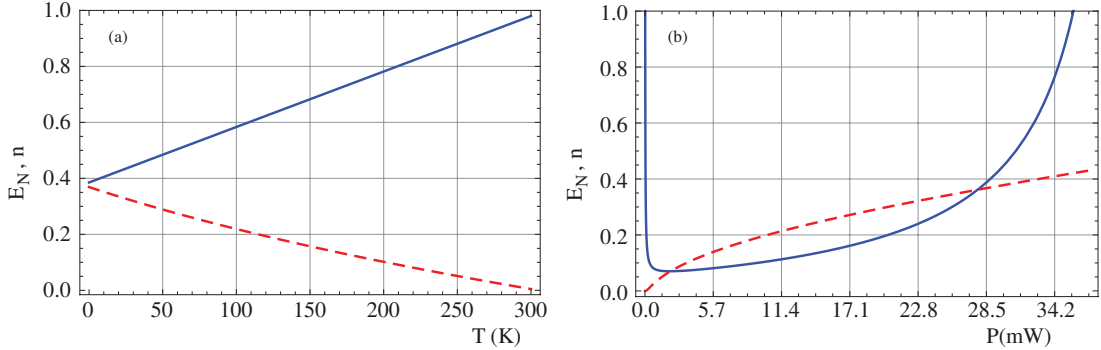}
  \caption{(a) Effective mean vibrational number $n$ (blue, full line) and logarithmic negativity  $E_{N}$ (red, dashed line) versus temperature, at fixed input power ${\mathcal P}=28.5$ mW; (b) versus input power ${\mathcal P}$, at fixed temperature $T=1$ K. Other parameters are $L = 0.74$ mm, $L_d= 50$ nm, $m=9$ ng, $\omega_m/2\pi=10$ MHz, $Q_m=4 \times 10^6$.}
\end{figure}

\begin{theacknowledgments}
This work has been supported by the European Commission (FP-7 FET-Open project MINOS), and by INFN (SQUALO project).
\end{theacknowledgments}

\bibliographystyle{aipproc}

\end{document}